\documentclass[12pt]{article}
\usepackage[cp1251]{inputenc}
\usepackage[T2A]{fontenc}
\usepackage{indentfirst}
\usepackage[dvips]{graphicx}
\begin{document}

                     Surprising attractive potential barriers and repulsive wells.

\vspace*{0.5cm}
               Zakhariev B.N.,

                Laboratory of theoretical Physics, Joint Institute for Nuclear Research,
               Dubna, Russia

\vspace*{1cm}                       Abstract

      We reveal the fundamental fact that in the old good quantum mechanics there is possible
      such unexpected inversion: potential barriers can drag in wave-particles and wells can push them off.

\vspace*{0.5cm}

             PACS:

\vspace*{1cm}
 This was hidden till now in the well-known convenient often used  multichannel formalism with matrix interactions
  and matrix wave functions and inaccessible inside "black boxes" of our mighty computers which perform for us
  cumbersome numerical calculations. It is especially interesting to find new properties which existence could not be
  supposed as seeming evidently  absolute nonsense.  Although usually the magic of scientific discoveries
  consist in more or less striking transition from unknown or “impossible” to something surprisingly  simple \cite{discover}.
  Such unexpected additions to our intuition are important for widening our basic notions for further search of
   surprises of nature. This is an important step to quantum control of complex systems. It opens the way to better
   understanding many-dimensional and many-body systems.

         The multichannel formalism is one of the most powerful tools to describe such objects by  reduction to solution
   of simple ordinary differential coupled Schr\"odinger equations. It becomes very popular after the famous papers by
 Feshbach \cite{Fesh}. We ourselves were then stimulated to generalize this practical apparatus to consider reactions
 with rearrangement of particles  published in by Feshbach's Ann. Phys. N.Y. \cite{EZZ} and later have even published
 a book \cite{ZhZ}.
 After that we have taken a part in development the quantum inverse problem multichannel formalism \cite{Less, multan},
 which  gave infinite number of exactly solvable models.

  The main step in the close coupled channels approach is the expansion of N-dimensional wave function
  $\Psi (x_{1},...,x_{N})$ in a set of known basic functions  $\Phi_{\alpha }(x_{2},...,x_{N})$ which depend on N-1
  variables \cite{EZZ,ZhZ}:
\begin{equation}
\Psi(x_{1},...,x_{N})=\sum_{\alpha } \Psi _{\alpha }(x_{1}) \Phi
_{\alpha }(x_{2},...,x_{N})
\label{expan}
\end{equation}
Here the coefficients $\Psi _{\alpha }(x_{1})$ are $\alpha $- channel functions (partial channel components).
They can be considered as a new vector representation of the initial wave function. Here $\alpha $
is a multi-index with so much components $(N-1)$ as the number
of variables in basic functions. This is like a substitution of part of $N-1$ continuous variables $x_{i\ge 2}$ by
the lattice of discrete channel variables $\alpha $. Here will be useful our experience in dealing with this
formalism \cite{multan}.

The expansion coefficients $\Psi _{\alpha }(x_{1})$ dependent on
one variable $x_{1}$ are found by solving the {\bf system} of
{\bf ordinary} differential equations (coupled one-dimensional
Schr\"odinger equations) instead of equations in partial derivatives
for $\Psi (x_{1},...,x_{N})$:
\begin{eqnarray}
-\frac{d^2}{dx^2} \Psi_{\alpha}(x)+ \sum_{\beta}V_{\alpha
\beta}(x)\Psi_{\beta}(x) =E_{\alpha}\Psi_{\alpha}(x), \enskip
E_{\alpha}=E-\epsilon_{\alpha },  \\
V_{\alpha \beta}(x)=\int d\xi \Phi_{\alpha }^{*}(\xi )V(x,\xi )
\Phi_{\beta }(\xi ).  \label{system}
\end{eqnarray}
Here we take $x \equiv x_{1}$, $\xi \equiv \{x_{i} \}_{i=2}^{N}$,
and $V(x,\xi )$ is the potential of corresponding
multi-dimensional problem. The eigenvalues $\epsilon_{\alpha }$
are related to basic functions $\Phi_{\alpha }$ and are threshold
energies above which the corresponding partial channels become open.
See more about the channel energies $E_{\alpha }$  in \cite{Less}.

The multichannel approach allows to describe the multi-dimensional and multi-particle
systems in nuclear, atomic, molecular physics, etc. It allows to
consider processes with excitation of their different internal degrees of
freedom and also to solve problems with mixing of spin
states, see \cite{Fesh,ZhZ,Less}.

The difference of the multi-channel equation (\ref{system}) from
{\it scalar} one-channel case consists in that here we have {\bf
vector} wave function $\Psi_{\alpha}(x)$ and instead of scalar potential
$V(x)$ we have interaction {\bf matrix} $V_{\alpha \beta}(x)$.
Also instead of scalar spectral weight factors (SWF) $c_{\nu }$
here serve spectral weight vectors (SWV) with components
$C_{\alpha m}= \Psi' _{\alpha}(0, E_{m})$ (in Gelfand-Levitan
approach and some other one in Marchenko formalism).

It is convenient to analyze  the system of one-dimensional equations (\ref{system})
comparing the rules for channel  wave  bending for the simplest   single-channel
case with  scalar  wave  $\Psi (x)$ instead of  vector valued one in (\ref{system}) under the
influence of  the given scalar  potential  $V(x)$  instead of matrix valued one:
\begin{eqnarray}
-\frac{d^2}{dx^2} \Psi_(x) = (E - V(x)) \Psi_(x)
\end{eqnarray}
 For positive kinetic energy $E-V(x)$  the simple  rule of bending $\Psi (x)$   is prescribed by the value of  its
 second derivative $\frac{d^2}{dx^2}\Psi (x)$ proportional to  $\Psi (x)$ itself . This rule is  to  curve the wave
{\bf to the  $x$-axis} independent on  sign of  $\Psi(x)$.
 It is illustrated by the limit unperturbed case $V(x)=0$ of free $\sin(\sqrt{E} x)$ wave. For negative $E-V(x)$ the
 bending  take place in opposite direction: {\bf from the $x$-axis}.

Taking this into account let us  get instructive generalized  bending rules  for the partial waves of multichannel
 formalism.  The diagonal terms $(E_{\alpha} - V_{\alpha \alpha }(x)) \Psi_{\alpha }(x)$  influence on the intensity
 $\frac{d^2}{dx^2}\Psi_{\alpha }(x)$ of bending $\Psi_{\alpha }(x)$ in analogy with the one-channel case: they
 yield to bending $\Psi (x)$
  {\bf to or from the $x$ axis}.
But the more extensive manifold of nondiagonal $V_{\alpha \beta }(x)$   appears to have  more diverse behavior.  For
the same signs of coupled channel
wave functions $\Psi _{\alpha }(x), \, \Psi_{\beta }(x)$  the interaction matrix $V_{\alpha \beta }(x)$ acts analogous to usual
  potential or diagonal matrix elements $V_{\alpha \alpha }(x)$ .  Their wells or  barriers increase or decrease the
  intensity $\frac{d^2}{dx^2}\Psi_{\alpha}(x)$ of  partial wave bending to $x$ axis. But for the opposite signs of
   $\Psi _{\alpha }(x), \, \Psi_{\beta }(x)$ there is paradoxically inverted situation: as if there are {\bf “attractive
barriers”! and “repulsive wells”!}. This unusual but nevertheless simple formulation of matrix interaction rules appears to
    be  useful for clear understanding some previously mysterious quantum peculiarities.
 Really for the same signs of $\Psi_{1}(x)$ and $\Psi_{1}(x)$
barriers in $V_{\alpha \alpha}(x)$ and $V_{\alpha \beta }(x)$ in both terms
 $-V_{\alpha \alpha }(x))\Psi_{\alpha }(x)$ and $-V_{\alpha \beta }(x) \Psi_{\beta }(x)$ in the $\alpha $ equation
 in \ref{system} are subtracted from the energy value $E_{\alpha }$ and make smaller the bending intensity
 |$\Psi"_{\alpha }(x)$|. Analogously act the potential matrix elements in the $\beta $ -equation in \ref{system}.
  Wells in $V_{\alpha \beta }(x)$ change|$\frac{d^2}{dx^2}\Psi_{\alpha }(x)$| and |$\frac{d^2}{dx^2}\Psi_{\beta }(x)$| also as diagonal matrix
  elements $V_{\alpha \alpha }(x)$ and $V_{\beta \beta }(x)$. But for different signs
 of  $\Psi_{\alpha }(x)$ and $\Psi_{\beta }(x)$ the {\it coupling terms in}
 (\ref{system}) {\it have the opposite influence}. It is better seen if we write the right hand side of partial equation
 in the form of {\it effective kinetic energy}  with common factor $\Psi_{\alpha }(x)$:
 $$E_{kin} \Psi_{\alpha}(x) \equiv [E_{\alpha} - V_{\alpha \alpha }(x)- V_{\alpha \beta }(x) \Psi_{\beta }(x)/\Psi_{\alpha }(x)]\Psi_{\alpha }(x)$$.
Here it is evident that the influence of the nondiagonal matrix element $V_{\alpha \beta }(x)$ directly depends on the
sign of the fraction $\Psi_{\beta }(x)/\Psi_{\alpha }(x)$, namely on relative signs of $\Psi_{\beta }$ and $\Psi_{\alpha }(x)$.
So trivial appears the rule for increasing or decreasing the {\bf effective kinetic energy} which straightforwardly
corresponds to attraction or repulsion of $V_{\alpha \beta }(x)$. Not only we get the simplest explanation of "impossible"
inversion: attraction $longleftrightarrow$ repulsion, but we get the universal rule for {\bf arbitrary} coupling term in
 {\bf any} interaction matrix.

  In conclusion we emphasize that the previously unmentioned peculiarity of nondiagonal elements of the interaction matrices
  is a very important correction of our quantum notions. This must facilitate the difficult task of understanding behavior
  of complex systems in future research.

Author is thankful to V.M. Chabanov for valuable collaboration.

\end{document}